\documentclass[prl,preprint,showpacs]{revtex4}                                                                            
\usepackage[sort&compress]{natbib}
\usepackage{amssymb}
\usepackage{amsbsy}             
\usepackage{bm}
\usepackage{mathrsfs}
\usepackage{graphics}
\usepackage{dcolumn}
\usepackage{graphicx}
\usepackage{amsmath}
\usepackage{amsthm}
\usepackage{epsfig}
\usepackage{latexsym}
\begin{document}
                                                                                                                         
\title{Time dependent density functional theory for nonadiabatic electronic dynamics.}
\addtocounter{page}{0}

                                                                                                                         
\author{Vinod Krishna}\email{vkrishna@hec.utah.edu}
\affiliation{University of Utah, Department of Chemistry, UT 84112}

\begin{abstract}  
  We show that the time dependent single electron, nuclear density matrix of an 
interacting electronic system coupled to nuclear degrees of freedom can be 
exactly reproduced by that of an electronic system with arbitrarily
specified electron-electron interactions coupled to the same nuclear 
degrees of freedom, given the initial density matrix of the interacting system. 
This formalism enables the construction of rigorous time dependent density functional 
theories to study nonadiabatic electronic dynamics. We obtain the Runge-Gross and 
Van Leeuwen theorems as special cases in the adiabatic limit. 
\end{abstract}
\maketitle
\pagenumbering{arabic}
 

 
 

\par  The correlated dynamics of electrons and nuclei drive fundamental mechanisms of 
chemical reaction dynamics. Most of these mechanisms involve nonadiabatic electron
nuclear dynamics, where the adiabatic or Born-Oppenheimer approximation is no longer 
valid.\cite{Butler,Wodtke} A central problem in understanding correlated electron-nuclear 
dynamics is the mutual time evolution of the electron and nuclear quantum subsystems. 
Nonadiabatic energy and information transfer between the two subsystems plays a fundamental 
role in correlated electron-nuclear dynamics. Nonadiabatic processes are controlled by nuclear 
momentum dependent derivative couplings between the electronic and nuclear subsystems. 
Because of their momentum dependence, these couplings also manifest themselves through 
complex geometric phase effects, a well known example of which is the Jahn-Teller effect
\cite{mead1,shapere1}. 
Nonadiabatic couplings between the two subsystems are also generally 
off diagonal in the nuclear subspace and introduce nonlocal, history dependent correlations 
into the electron-nuclear dynamics. 
\par Due to the complexity of the time evolution, developing $\it{ab}$ $\it{initio}$ 
methods to model correlated electron-nuclear dynamics has proved to be a challenging problem. 
Time dependent density functional theory(TDDFT) is a formally exact, successful method 
developed to understand the dynamical properties of interacting electron systems. This 
theory relies on the Runge-Gross theorem \cite{Runge} and its generalization by Van 
Leeuwen \cite{Van}, which relate the exact single electron density to the single 
electron time dependent potential. It is also a computationally tractable method, and 
hence is a promising approach that could be extended to model correlated electron-nuclear 
dynamics. TDDFT has recently been extended to model certain open quantum systems \cite{burke,diventra}.  
\par Direct extensions of TDDFT to model correlated electron-nuclear dynamics involve the 
estimation of only single particle densities and more generally diagonal elements of the 
system density matrix \cite{butriy,kreibich1,kreibich} and are based on extensions of DFT to 
multicomponent systems \cite{tong}. Although they guarantee the correct estimation of 
these quantities, they are unable to fully account for the dynamical phase correlations due 
to the nonadiabatic derivative couplings between electronic and nuclear subsystems. These 
correlations are of fundamental importance to nonadiabatic energy transfer mechanisms that 
occur in many chemical reaction processes \cite{Butler}. This is because nonadiabatic phase 
correlations that arise during the system's time evolution cannot be fully captured by 
diagonal density matrix elements. It is also difficult to derive essential simplifying 
semiclassical approximations to the multicomponent TDDFT framework that are required 
for tractable computational studies.        
   We propose a formulation that extends the density functional paradigm to model the time 
evolution of interacting electron nuclear systems starting from an arbitrary initial density 
matrix for the entire system. The theory we present is designed to explicitly and exactly 
capture nonadiabatic effects on correlated electron-nuclear dynamics while simplifying the
effects of electron-electron interactions. Our theory also retains off diagonal elements
of the density matrix in the nuclear subspace. This property allows our theory to retain 
information regarding quantum coherences during the system time evolution. Thus, in contrast
to the multicomponent DFT approach of Refs.(9)-(11), our theory enables an accurate treatment 
of dynamical phenomena induced by nonadiabatic transitions during electron-nuclear dynamics. 
This formalism also allows for the straightforward construction of semiclassical approximations 
to the nuclear dynamics, and also separably includes the effects of nonadiabatic couplings, 
thus allowing for approximations to model nonadiabatic electronic dynamics based on traditional TDDFT. 
   To construct this theory, we map the interacting electronic system into a 
reference system where electron-electron interactions can be specified according to 
convenience, coupled to nuclear degrees of freedom whose interactions are retained.
Let $H({\bf q, R})$ be the combined Hamiltonian of the system, with $\{{\bf R,P}\}$
labeling the nuclear coordinates and $\{{\bf q}\}$ the electronic coordinates:
\begin{equation}
   H({\bf q, R})  = \frac{P^{2}}{2M} + H_{0}({\bf q}) + V({\bf q};{\bf R},t)
\end{equation}
  $V({\bf q};{\bf R},t)$ is the sum of the time dependent single electron-nuclear 
potential and the purely nuclear-nuclear interaction in the Hamiltonian. 
  The density matrix, $\hat{\rho}(t)$, of the system evolves according to the 
Liouville equation:
\begin{equation}
i\hbar\frac{\partial}{\partial t}\hat{\rho}(t) = \big[H,\hat{\rho}(t)\big]
\end{equation}
  A key issue is that a theory for nonadiabatic dynamics should be able to evaluate 
time dependent properties of electronic dynamics while simultaneously retaining information 
regarding dynamics in the nuclear subspace. A natural physical variable that contains this 
information is the single electron, reduced nuclear density matrix which is conjugate to 
the electron-nuclear potential. This matrix is made up of diagonal contributions from the 
electronic subspace, corresponding to the electronic probability density, while the nuclear 
part of the density matrix is fully retained. 
  We derive continuity equations for reduced electron-nuclear density and current 
matrices and relate their time evolution directly to the electron-nuclear 
potential,$V({\bf q};{\bf R},t)$. A suitable and physically transparent method to 
derive the continuity equations is obtained from analyzing partial Wigner transforms 
of the density matrix in the nuclear subspace:
\begin{equation}
\tilde{\rho}({\bf R,P};t) = \int{d{\bf z}\exp[-\frac{i\bf P\cdot\bf z}{\hbar}]
\langle{\bf R}+\frac{\bf z}{2}\vert\hat{\rho}(t)\vert{\bf R}-\frac{\bf z}{2}\rangle}
\end{equation}
 The quantum Liouville equation can be written in terms of the partial Wigner
transformed density matrix:
\begin{equation}
i\hbar\frac{\partial}{\partial t}\tilde{\rho}({\bf R,P};t) = 
\left\{\tilde{H}({\bf R,P}),\tilde{\rho}({\bf R,P};t)\right\} 
\end{equation}
The bracket $\left\{..\right\}$ is the Wigner-Weyl-Moyal bracket
 \cite{Wig,Moy,Groen}. The bracket between two arbitrary 
operators $A$ and $B$ is defined as:
\begin{align}
\left\{A, B\right\} =\tilde{A}({\bf R,P})*\tilde{B}({\bf R,P}) - 
\tilde{B}({\bf R,P})*\tilde{A}({\bf R,P}) 
\end{align}
The Moyal product,'$*$', is defined\cite{Moy} as a bilinear product of the Wigner 
transforms of $\hat{A}$ and $\hat{B}$:
\begin{align}
\tilde{A}({\bf R, P})*\tilde{B}({\bf R,P})= 
\tilde{A}({\bf R, P})\exp\Big[\frac{\hbar\Lambda}{2i}\Big]\tilde{B}({\bf R,P}) &\nonumber\\
\Lambda({\bf R, P})=\overleftarrow{\nabla}_{{\bf P}}\cdot\overrightarrow{\nabla}_{{\bf R}}-
\overleftarrow{\nabla}_{{\bf P}}\cdot\overrightarrow{\nabla}_{{\bf R}}&
\end{align}
The Liouville equation for the partial Wigner transformed density, 
$\tilde{\rho}\equiv\tilde{\rho}({\bf R,P};t)$, is:
\begin{align}
i\hbar\frac{\partial\tilde{\rho}}{\partial t} =
\left\{\frac{P^{2}}{2M} + H_{0}({\bf q}),\tilde{\rho}(t)\right\} + 
&\left\{\tilde{h}({\bf q;R}),\tilde{\rho}\right\}
\end{align}
Writing $\Re\equiv\{{\bf R,P}\}$, single particle electronic variables are
defined as traces over $\tilde{\rho}(\Re;t)$:
\begin{equation}
f({\bf x}\vert\Re,t) = Tr_{el}\left\{\hat{f}({\bf x})\tilde{\rho}(t)\right\}
\; ;\;\hat{f}\equiv \{n,\vec{j}\} 
\end{equation}
The current and density operators are:
\begin{align}
\hat{n}({\bf x}) \equiv \frac{1}{N}\sum_{i=1}^{N}{\delta({\bf x}-\hat{\bf q}_{i})}\\
\hat{\vec{j}}({\bf x}) \equiv \frac{1}{2m}\sum_{i=1}^{N}
{\{\hat{\bf p}_{i},\delta({\bf x} -\hat{\bf q}_{i})\}} 
\end{align}
We also define the quantities:
\begin{align}
\vec{\gamma}_{v}({\bf x}\vert\Re ;t) \equiv  
Tr_{el}\left\{\big[\hat{\vec{j}}({\bf x}),H_{0}({\bf q})\big]
\hat{\rho}_{W}(\Re;t)\right\} \nonumber\\
i\hat{L}[v](\Re,t) \equiv \frac{\bf P}{M}\cdot\nabla_{\bf R} + 
\frac{i}{\hbar}\left\{v({\bf x;R},t), \right\}
\end{align}
From these definitions, the equation of motion for the single electron 
density and current are obtained: 
\begin{align}
\big[\partial_{t} + i\hat{L}[v](\Re,t)\big]n({\bf x}\vert\Re,t) = 
-\nabla_{x}\cdot\vec{j}({\bf x}\vert\Re,t)\\
\big[\partial_{t} + i\hat{L}[v](\Re,t)\big]\vec{j}({\bf x}\vert\Re,t) = 
-\vec{\gamma}_{v}({\bf x}\vert\Re,t)\nonumber\\
-\frac{1}{m}\nabla_{\bf x}v({\bf x;R},t)*n({\bf x}\vert\Re,t)
\end{align}
  In the adiabatic limit, Eqs.(12) and (13) reduce to the traditional continuity 
and current equations for a system of interacting electrons in an external time 
dependent potential. However, Eq.(12) and Eq.(13) are more general and include 
the effect of electron-nuclear nonadiabatic couplings. This is given by the 
action of the quantum Liouville operator, $i\hat{L}[v]$ on the single electron 
distribution functions, $\{n,\vec{j}\}$. If the classical limit is taken 
for the nuclei alone, then the time evolution of the electron density has 
two contributions, one corresponding to the usual continuity equation in 
the electronic subspace, and the second from the time evolution of a classical 
nuclear subspace off which electrons can scatter. 
  A satisfactory reference system would need to reproduce correctly the 
electronic density $n({\bf x}\vert\Re,t)$. We show that the single electron density 
matrix and its corresponding current density matrix can be reproduced by several 
hamiltonians which differ from the original hamiltonian in the strength of their 
electron-electron correlation, through the construction of an appropriate single 
electron-nuclear interaction for each such hamiltonian. 
\newline
$\it{Statement-}$Let $H({\bf q;R})$ be a Hamiltonian with a single particle 
electron-nuclear coupling and a electron-electron interaction potential given by 
the pair $\{v({\bf x;R,t}),V_{ee}({\bf r})\}$.Given an initial value, 
$\{n({\bf x}\vert\Re,0),\vec{j}({\bf x}\vert\Re,0)\}$, and a pair of solutions, 
$\{n({\bf x}\vert\Re,t),\vec{j}({\bf x}\vert\Re,t)\}$ to the current and continuity 
equations, Eqs.(12) and (13), a second Hamiltonian $H'$ with an arbitrarily 
specified electron-electron interaction, $W_{ee}({\bf r})$, can be  
constructed to reproduce the solutions, $\{n({\bf x}\vert\Re,t),\vec{j}({\bf x}
\vert\Re,t)\}$ by modifying its single electron-nuclear potential, $w({\bf r}
\vert\Re;t)$. 
\newline
$\it{Proof-}$We present a heuristic and constructive derivation. Assume that there exists 
a Hamiltonian, $H'$ with a pair of interactions, $\{w({\bf x;R}),W_{ee}({\bf q})\}$ 
from which the single particle density $n({\bf x}\vert\Re;t)$, and current, 
$\vec{j}({\bf x}\vert\Re;t)$ can be derived. We will show that for an arbitrary 
two particle electron-electron interaction, $W_{ee}$, a single particle electron-nuclear 
coupling $w$ can be constructed that reproduces the given electron density. By assumption, 
Eq.(12) and Eq.(13) are satisfied for both the Hamiltonians, $H({\bf q;R})$
and $H'({\bf q;R})$. For notational convenience, we define $\zeta({\bf x}\vert\Re;t) \equiv
v({\bf x}\vert\Re;t) - w({\bf x}\vert\Re;t)$ and $\vec{\gamma}_{vw} = \vec{\gamma}_{v} 
-\vec{\gamma}_{w}$. Then by a process of subtraction, we find that the single 
particle current and density satisfy:
\begin{eqnarray}
\vec{\gamma}_{wv}({\bf x}\vert\Re;t) =
\frac{\nabla_{\bf x}\zeta}{m}*n({\bf x}\vert\Re;t)
 + \frac{i}{\hbar}\left\{\zeta,\vec{j}({\bf x}\vert\Re;t)\right\}\\ 
\left\{\zeta({\bf x}\vert\Re,t), n({\bf x}\vert\Re,t)\right\} = 0
\end{eqnarray}
  Two operators $A,B$ star commute, i.e the equation, $A*B - B*A = 0$ is satisfied 
when $A$ is of the form, $A=g({\bf x},t);$ or $A=F(B,*)$, where $F(B,*)$ is a 
function generated from Moyal star product polynomials in $B$, i.e. 
$A=\sum_{k}{g_{k}({\bf x},t)B*B*B\ldots\text{k times}}$. In addition to these solutions, 
other functions could exist whose star product commutator with $B$ is zero. 
Thus, the solutions to Eq.(15) can be written in the form:
\begin{equation} 
\zeta({\bf x}\vert\Re;t) = \zeta[{\bf x},t;n({\bf x}\vert\Re;t),\Omega(\Re;t)]
\end{equation}
  Here, $\Omega(t)$ contains the set of functions which star commute with $n$, and
which cannot be generated from the single particle density $n$. In analogy with 
quantum mechanical language, we can say that the set of functions, $n,\Omega$ form 
a complete set of commuting variables. 
 To understand the solutions to Eq.(14), we first study it in the approximation 
where the classical limit is taken for the nuclear variables. In this limit Eq.(14) 
becomes:
\begin{align}
\frac{1}{m}n\nabla_{\bf x}\zeta+
\left\{\zeta({\bf x}\vert\Re;t),\vec{j}({\bf x}\vert\Re;t)\right\}_{PB} 
= \vec{\gamma}_{wv}^{cl}({\bf x}\vert\Re;t)
\end{align}
Here the bracket, $\{..\}_{PB} \equiv \overleftarrow{\nabla}_{\bf R}\cdot
\overrightarrow{\nabla}_{\bf P} - \overleftarrow{\nabla}_{\bf P}\cdot
\overrightarrow{\nabla}_{\bf R}$ is the classical Poisson bracket defined 
over the nuclear subspace. Physically, the first term of Eq.(17) is the 
electronic force due to the change in the single particle potential, 
while the second term is the nonadiabatic coupling contributions that are a 
result of varying the electron-nuclear coupling. These two terms add to balance 
the force contributions due to purely electron-electron correlations. 
In the absence of the nonadiabatic coupling between electrons and nuclei, 
the equation reduces to the force acting on the electrons due to the
difference in single particle potentials, which exactly compensates for the 
difference in electron-electron interactions, as in traditional TDDFT. The 
nonadiabatic coupling adds a new contribution corresponding to exchange of 
energy between electronic and nuclear degrees of freedom. 
We show below that if appropriate spatial boundary conditions are specified,
Eq.(14) can be solved in the mixed quantum classical limit (Eq.(17)) at a 
given time, $t=0$ and these solutions can be propagated to succeeding timesteps 
in a iterative fashion. To construct an explicit solution of Eq.(17), we define:
\begin{equation}
\{\vec{v}_{\alpha},\vec{u}_{\alpha}\} \equiv \{\nabla_{P_{\alpha}}\vec{j},
\nabla_{R_{\alpha}}\vec{j}\}; \alpha = x,y,z. 
\end{equation}
  Eq.(17) can be rewritten as:
\begin{align}
\nabla_{\bf x}[n({\bf x}\mid\Re;t)\zeta] + \sum_{\alpha}{\big[\vec{v}_{\alpha}\nabla_{R_{\alpha}} - 
\vec{u}_{\alpha}\nabla_{P_{\alpha}} - \nabla_{\bf x}n\big]}\zeta
= \vec{\gamma}_{wv}({\bf x}\mid\Re;t) \hspace{1.0cm}
\end{align}
  We define the operator $\hat{\Lambda}$ and a normalized potential $\sigma$ as:
\begin{align}
i\Lambda({\bf x}\vert\Re;t) =\Big[\sum_{\alpha}{\big[\vec{v}_{\alpha}\nabla_{R_{\alpha}}\frac{1}{n} -
\vec{u}_{\alpha}\nabla_{P_{\alpha}}\frac{1}{n}\big]} - \frac{\nabla_{\bf x}n}{n}\Big]\nonumber\\
\sigma({\bf x}\vert\Re;t) = n({\bf x}\vert\Re;t)\zeta({\bf x}\vert\Re;t) \hspace{0.5cm}
\end{align}
We also assume that the ratios $\nabla_{..}\vec{j}/n$ and $\nabla_{..}n/n$ vanish as 
${\bf x}\rightarrow\ \pm\infty$. Eq.(17) becomes:
\begin{equation}
\nabla_{\bf x}\sigma + i\vec{\Lambda}\sigma = \vec{\gamma}_{wv}^{cl}
\end{equation} 
This equation can be solved explicitly as follows:
  Define the operator, $\hat{\Gamma}$ such that it satisfies:
\begin{equation}
\nabla_{\bf x}\hat{\Gamma} = i\hat{\Gamma}\vec{\Lambda}({\bf x}\mid\Re;t) 
\end{equation}
Then $\hat{\Gamma}$ is a path ordered exponential in the electronic subspace given by:
\begin{equation}
\hat{\Gamma}({\bf x}\mid\Re;t) \equiv \int{d{\bf y}\sum_{\mathscr{C}:{\bf y\rightarrow\bf x}}{{\mathscr P}
\exp{\Big[i\int^{\bf x}_{\bf y}{\vec{\Lambda}({\bf z}\vert\Re;t)\cdot d{\bf z}}\Big]}}}
\end{equation}
 ${\mathscr P}\exp{[..]}$ in Eq.(23) is the path ordering exponential operator. The operator 
$\hat{\Gamma}$ is a path integral over all paths, $\mathscr{C}:{\bf y\rightarrow x}$, ending in 
$({\bf x},t)$ in the electronic subspace. Since this is a symmetric operator it has an inverse, 
$\hat{\Gamma}^{-1}$. 
As a consequence of the boundary conditions, $\Gamma$ becomes unity as ${\bf x}\rightarrow\pm\infty$. 
Eq.(20) is rewritten as:
\begin{equation}
\hat{\Gamma}^{-1}\nabla_{\bf x}\hat{\Gamma}\sigma = \vec{\gamma}_{wv}^{cl}
\end{equation}
 This can be solved to obtain:
\begin{equation}
\sigma = \hat{\Gamma}^{-1}({\bf x}\mid\Re;t)
\int^{\bf x}{\hat{\Gamma}({\bf y}\mid\Re;t)\vec{\gamma}_{wv}^{cl}({\bf y}\mid\Re;t)\cdot d{\bf y}}
\end{equation}
Thus, a formal solution for Eq.(17) has been constructed. By requantizing the classical 
nuclear variables using the Weyl correspondence, a solution can be generated for the fully 
quantum dynamics, Eq.(14). The solution also has a transparent physical interpretation. The 
kernel, $\Gamma^{-1}({\bf x})\Gamma({\bf y})$ contains the "history" of the interactions between  
the electronic and nuclear degrees of freedom. In the Born-Oppenheimer limit, the path ordering 
becomes unneccesary and the kernel becomes independent of path. Thus, this kernel builds 
into the effective potential, electronic scattering and nonadiabatic dynamical 
phase correlations.  
\par  We now describe the time propagation of the solutions to Eq.(14) using a method similar
to that developed in Ref. \cite{Van}. We require as an initial condition, that the initial 
density matrix of the system is known. Consequently, the quantity, $\vec{\gamma}({\bf x}\vert\Re;0)$ 
and its time derivatives at $t=0$ can be evaluated from the initial value of the 
density matrix $\hat{\rho}_{W}(\Re;0)$. Furthermore we make the assumption that all 
the relevant potentials and distribution functions are Taylor expandable around the 
time $t=0$. The Taylor expansions for such functions $f({\bf x}\vert\Re;t)$ are 
defined by:
\begin{equation}
f({\bf x}\vert\Re;t) = \sum_{m=0}^{\infty}{f^{(m)}(0)\frac{t^{m}}{m!}}
\end{equation}
Using this Taylor expansion, Eq.(14) can be rewritten as a system of 
linear difference equations: 
\begin{align}
\vec{\gamma}_{wv}^{(k)}(0) = \sum_{l=0}^{k}{C^{k}_{l}\Big[\frac{1}{m}\nabla_{x}
\zeta^{(l)}(0)*n^{(k-l)}(0)}
+ \frac{i}{\hbar}\left\{\zeta^{(l)}(0),\vec{j}^{(k-l)}(0)\right\}\Big] 
\end{align}
This equation can be solved iteratively. For $k=0$ and $k=1$, the system of 
equations become:
\begin{align}
\vec{\gamma}_{wv}^{(0)}(0) =\frac{1}{m}\nabla\zeta^{(0)}(0)*n^{(0)}(0) + 
\frac{i}{\hbar}\left\{\zeta^{(0)},\vec{j}^{(0)}\right\}
\end{align}
\begin{align}
\vec{\gamma}_{wv}^{(1)}(0) = \frac{1}{m}\nabla\zeta^{(1)}(0)*n^{(0)}(0) + 
\frac{1}{m}\nabla\zeta^{(0)}(0)*n^{(1)}(0)
+ \frac{i}{\hbar}\Big[\left\{\zeta^{(1)},\vec{j}^{(0)}\right\}
+\left\{\zeta^{(0)},\vec{j}^{(1)}\right\}\Big]\hspace{0.5cm}
\end{align}
Eq.(27) has one unknown, $\zeta^{(0)}$, and all the other quantities are known, 
given the assumptions made. It has a unique solution, upto a function of only the 
nuclear coordinates whose Wigner-Moyal bracket with the current $\vec{j}^{(0)}(0)$ 
is zero. Furthermore, the second equation also has a solution with similar properties, 
given $\zeta^{(0)}(0)$, since it also satisfies an equation of the same mathematical 
form. This argument can be inductively extended to solve for the $k$-th time 
derivative$\zeta^{k}(0)$, for an arbitrary value of $k$. Thus, this method 
provides a constructive time dependent solution to Eq.(14). 
\par We now consider an important special case. When the electronic dynamics of the 
system is adiabatic, the timescales on which the electronic dynamics occur are much
faster than the timescales for nuclear motion. As a result, the time derivative of the 
single electron-nuclear density matrix is dominated by electronic dynamics, and the 
derivatives w.r.t nuclear degrees of freedom can be neglected. This implies that the 
continuity and current equations become:
\begin{align}
\partial_{t}n({\bf x}\vert\Re;t) \approx -\nabla_{\bf x}\vec{j}({\bf x}\vert\Re;t)\\
\partial_{t}\vec{j}({\bf x}\vert\Re;t) \approx -\vec{\gamma}_{v}({\bf x}\vert\Re;t) -
\frac{1}{m}n({\bf x}\vert\Re;t)\nabla_{\bf x}v
\end{align}
  Eqs.(30) and (31) are of the same form as the equations obtained for a purely electronic
time evolution under an external time varying potential, $v$. The Runge-Gross theorem \cite{Runge} 
has been generalized to relate the time evolution of a electron density and current $\{n,\vec{j}\}$ 
to an external potential for systems that satisfy equations of this form \cite{Van}. Thus, in the 
adiabatic limit, Van Leeuwen's generalization of the Runge-Gross theorem is obtained from this formalism.  
  In conclusion, we have shown that the Runge-Gross theorem can be generalized
to describe the correlated nonadiabatic dynamics of an interacting electron-nuclear
system. This theory is very general and allows maximal information about the quantum 
mechanical nuclear variables to be retained. Numerically tractable DFTs can be constructed 
from the explicit solutions, Eq.(25),by the application of various semiclassical reduction 
schemes on the nuclear dynamics. For example, for weakly nonadiabatic mixed quantum classical 
evolution, the nonadiabatic kernels in Eq.(25) can be evaluated using a Monte Carlo approach 
with the nonadiabatic derivative operators in Eq. (22) being treated as inducing occassional
''nonadiabatic transitions" according to a suitable Monte Carlo criterion. The construction of 
such functionals will be discussed in a future work. This formalism demonstrates that the density functionals 
required to correctly approximate electronic nonadiabatic dynamics are nonlocal functionals  
and depend on derivatives in the nuclear position and momentum variables. The nonlocality
is reminiscent of the memory dependence found in standard TDDFT \cite{maitra}. The effective 
potentials include quantum mechanical coherence and geometric phase factors that are a consequence 
of the nonadiabatic coupling between nuclear and electronic degrees of freedom.    
\newline I thank M. DiVentra, K. Burke, N. T. Maitra and A. V. Madhav for useful comments. 
\nonumber

\end{document}